\documentclass[runningheads]{llncs}

\usepackage{float}
\usepackage{amsmath}
\usepackage{amsfonts}
\usepackage{amssymb}
\usepackage{mathtools}
\usepackage{listings}
\usepackage{xcolor} 
\usepackage{todonotes}
\usepackage{hyperref}
\usepackage{bsymb,b2latex}
\usepackage[ruled, vlined, linesnumbered]{algorithm2e}
\usepackage{caption}
\usepackage{cleveref}
\usepackage{wrapfig}
\usepackage{tablefootnote}
\usepackage{tikz}

\usepackage{amsthm}

\newtheorem*{example*}{Example}
\newtheorem*{proposition*}{Proposition}

\crefname{lstlisting}{listing}{listings}
\Crefname{lstlisting}{Listing}{Listings}

\crefname{algocf}{alg.}{algs.}
\Crefname{algocf}{Algorithm}{Algorithms}

\definecolor{orcidlogocol}{HTML}{A6CE39}
\usepackage{tikz}
\usetikzlibrary{svg.path}
\tikzset{
  orcidlogo/.pic={
    \fill[orcidlogocol] svg{M256,128c0,70.7-57.3,128-128,128C57.3,256,0,198.7,0,128C0,57.3,57.3,0,128,0C198.7,0,256,57.3,256,128z};
    \fill[white] svg{M86.3,186.2H70.9V79.1h15.4v48.4V186.2z}
                 svg{M108.9,79.1h41.6c39.6,0,57,28.3,57,53.6c0,27.5-21.5,53.6-56.8,53.6h-41.8V79.1z M124.3,172.4h24.5c34.9,0,42.9-26.5,42.9-39.7c0-21.5-13.7-39.7-43.7-39.7h-23.7V172.4z}
                 svg{M88.7,56.8c0,5.5-4.5,10.1-10.1,10.1c-5.6,0-10.1-4.6-10.1-10.1c0-5.6,4.5-10.1,10.1-10.1C84.2,46.7,88.7,51.3,88.7,56.8z};
  }
}

\renewcommand{\orcidID}[1]{%
  \resizebox{8px}{8px}{
      \href{https://orcid.org/#1}{\tikz[yscale=-1,transform shape]{\pic{orcidlogo}}}}%
}

\begin{document}
\title{Failure divergence refinement for Event-B\thanks{The research presented in this paper has been partially conducted within the IVOIRE project, which is funded by ``Deutsche Forschungsgemeinschaft'' (DFG) and the Austrian Science Fund (FWF) grant \# I 4744-N.}}
%
%
 \author{
 Sebastian Stock\inst{1}\orcidID{0000-0002-2231-8656} \and
 Michael Leuschel\inst{2}\orcidID{0000-0002-4595-1518}  \and
 Atif Mashkoor\inst{1}\orcidID{0000-0003-1210-5953} 
 }

%
%
 \institute{
 Institute for Software Systems Engineering, Johannes Kepler University Linz\inst{1}\\
\email{\{firstname.lastname\}@jku.at} \\
 Institut f\"{u}r Informatik, Universit\"{a}t D\"{u}sseldorf\inst{2}\\
\email{leuschel@uni-duesseldorf.de }
}

\maketitle              
\begin{abstract}
When validating formal models, sizable effort goes into ensuring two types of properties: safety properties (nothing bad happens) and liveness properties (something good occurs eventually. Event-B supports checking safety properties all through the refinement chain. The same is not valid for liveness properties. Liveness properties are commonly validated with additional techniques like animation, and results do not transfer quickly, leading to re-doing the validation process at every refinement stage.
This paper promotes early validation by providing failure divergence refinement semantics for Event-B. We show that failure divergence refinement preserves trace properties, which comprise many liveness properties, under certain natural conditions.
Consequently, re-validation of those properties becomes unnecessary. Our result benefits data refinements, where no abstract behavior should be removed during refinement. Furthermore, we lay out an algorithm and provide a tool for automatic failure divergence refinement checking, significantly decreasing the modeler's workload. The tool is compared and evaluated in the context of sizable case studies.
\end{abstract}
\keywords{Refinement, Failure traces, Event-B, Divergence, Liveness}

\section{Introduction}
\label{sec:introduction}
Formal models are used to ensure the consistency 
 of requirements.
While creating a formal model, the modeler is concerned with two fundamental tasks: verification and validation. Verification checks for internal consistency, e.g., the preservation of invariants and the well-definedness of expressions.
Validation checks whether the desired behavior 
is part of the model.  
For validation, we can distinguish between showing the absence of bad states, i.e., safety properties, and the presence of desirable behavior, i.e., liveness properties~\cite{Roscoe1998}.

Event-B's~\cite{Abrial2010} standard proof obligations (PO) are aimed at verifying safety properties even during refinement, as pointed out by Hoang and Abrial~\cite{Hoang2011}.
Validation obligations (VO)~\cite{mashkoor21a}, on the other hand, provide support for validation. However, POs and VOs face challenges for checking liveness properties.

The drawback is significant: Assumptions about the abstract model's behavior do not automatically apply to the refining model. Consequently, the modeler confronts a dilemma: either redo sizable parts of the validation work for each refinement step, which can be time-consuming, or defer complete validation until the model is more concrete, risking project integrity if errors are too late.

We introduce a technique to address these issues by establishing failure divergence refinement for Event-B models. This approach extends standard Event-B refinement, enabling reasoning about preserving traces, a part of liveness properties.
Failure divergence refinement ensures that every abstract trace corresponds to a concrete trace, keeping the representation of desirable traces.
Once a failure divergence refinement relationship is established, trace-based validation results from a refined Event-B model can be applied to the refining model without requiring re-validation, thereby promoting early validation. 
It is thus also helpful to check that a data refinement does not accidentally remove behavior.
By providing a fully automatic tool within the ProB~\cite{Leuschel2003} environment, we make this process fully automatic and avoid the complexity of adding new proof obligations (POs). We can also provide the modeler with counterexamples where no refinement relationship can be established. Thus, we additionally enable targeted debugging of faulty refinement relationships.

The paper proceeds as follows: In \Cref{sec:background}, we provide an overview of Event-B, its refinement calculus, and the notion of traces. \Cref{sec:traces_failure} will tailor the known notions of trace refinement, failure traces, divergence, and failure divergence refinement to Event-B. \Cref{sec:formalization} then provides the first main contribution by showing that, given a natural condition, traces are preserved if two machines are in failure divergence refinement to each other. \Cref{sec:approach} discusses our second contribution, an implementation to check Event-B machines for failure divergence refinement. \Cref{sec:evaluation} presents an evaluation of this implementation on existing case studies. We explore the complementarity of our approach with existing techniques in \Cref{sec:related} and conclude with future directions in \Cref{sec:conclusion}.

\section{Background}
\label{sec:background}
\textbf{Event-B} is a formal modeling language based on set theory and first-order logic. Modelers create so-called \texttt{machines}, which can be seen as state machines. Typically, such machines are created and maintained with the help of the Rodin~\cite{Abrial2010a} platform, which helps discharge and track POs. 
\Cref{lst:vending} shows a basic example of a vending machine. There are two \texttt{variables}\footnote{In practice, an \texttt{invariant} clause defines the variables, and many POs rely on these invariants. However, we assume the presence of invariants implicit as indicated by inlining them in the \texttt{variables} clause. For the rest of this paper, we solely focus on the state space induced by variables and invariants, assuming there are no invariant violations, which is enforced via POs.}, \texttt{stock} and \texttt{coin}, and with the help of the \texttt{events}, we can insert coins, vend some beverages, or restock the system if it runs empty. 

\subsubsection*{Refinement}
\label{subsec:eventb_ref}
Refinement in Event-B is a way to extend an existing machine rigorously. Refining an abstract machine aims to make it more concrete by gradually adding complexity\footnote{A detailed overview is given by Hoang~\cite{Hoang2013}.}. One or more concrete events can refine abstract events.
Furthermore, the modeler can introduce new events, refining the virtual \texttt{skip} event, which is invisible at the abstract level.
The Rodin tool generates additional POs to ensure a correct refinement.

A refinement for \Cref{lst:vending} can be seen in \Cref{lst:vending2}, where we partitioned
the generic \texttt{stock} of beverages into specific stocks for \texttt{soda} and \texttt{water}.
As previously discussed, concrete \texttt{events} can refine abstract events in various ways. \texttt{insert\_coin} is refined by just one event. The \texttt{vend} event is refined by \texttt{vend\_soda} and \texttt{vend\_water}. \texttt{select\_drink} refines the invisible abstract event \texttt{skip}.

\lstset{basicstyle=\ttfamily\color{black}\scriptsize,
float,
captionpos = b}

\begin{figure}
\noindent\begin{minipage}{.47\textwidth}
\begin{lstlisting}[mathescape, label={lst:vending},  caption={Vending machine}, captionpos=b, numbers=left, stepnumber=1, frame=single,framexleftmargin=2.5em, framexrightmargin=-2.5em]
$\mathbf{machine}$ m0
$\mathbf{variables}$
 stock $\in \mathbb{N}$ $\land$ coin $\in \mathbb{N}$

$\mathbf{events}$
  INITIALISATION $\triangleq$ $\mathbf{then}$
    stock :=  3
    coin := 0
  $\mathbf{end}$

  insert_coin $\triangleq$  $\mathbf{when}$
  stock > 0 $\land$ coin + 1 $\leq$ stock
  $\mathbf{then}$
   coin := coin + 1
  $\mathbf{end}$

  vend $\triangleq$  $\mathbf{when}$
   coin > 0 $\land$ stock > 0
  $\mathbf{then}$
   stock :=  stock - 1
   coin := coin - 1
  $\mathbf{end}$

  restock $\triangleq$ $\mathbf{when}$
   stock = 0
  $\mathbf{then}$
   stock := 3
  $\mathbf{end}$
end
\end{lstlisting}
     \includegraphics[width=1\textwidth]{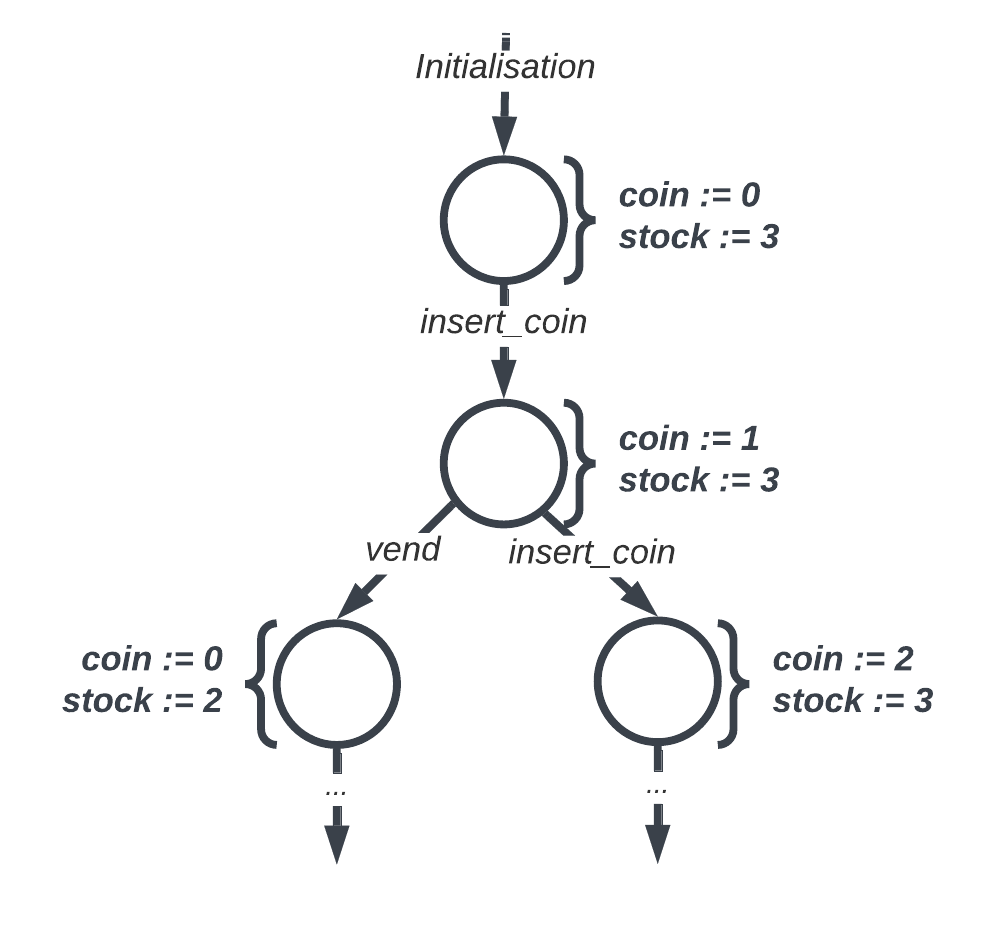}
    \caption{Visualization of the state space of \Cref{lst:vending}}
    \label{fig:trace1}
\end{minipage}\hfill %
\begin{minipage}{.50\textwidth}
\begin{lstlisting}[mathescape, label={lst:vending2}, caption={Vending machine refining \Cref{lst:vending}}, captionpos=b, numbers=left, stepnumber=1, frame=single,framexleftmargin=2.2em, framexrightmargin=-1.5em]
$\mathbf{machine}$ m1 $\mathbf{refines}$ m0 
$\mathbf{variables}$ 
soda $\in \mathbb{N}$ $\land$ water $\in \mathbb{N}$ $\land$
soda + water = stock $\land$ 
selectedDrink $\in$ {soda, water, none}

$\mathbf{events}$
  INITIALISATION  $\mathbf{then}$
   soda := 2 || water := 1 || 
   coin := 0 || selectedDrink := none
  $\mathbf{end}$

  $\mathbf{insert\_coin}$ $\mathbf{refines}$ $\mathbf{insert\_coin}$ $\triangleq$  $\mathbf{when}$
   soda + water > 0 $\land$
   coin + 1 $\leq$ soda + water $\land$
   selectedDrink = none
  $\mathbf{then}$
   coin := coin + 1
  $\mathbf{end}$

  $\mathbf{select\_drink} \triangleq$  $\mathbf{any}$ drink
  $\mathbf{where}$
   selectedDrink = none $\land $ coin > 0 $\land$ 
   drink $\in$ {soda, water} $\land$ 
   (drink = soda $\limp$ soda > 0) $\land$ 
   (drink = water $\limp$ water > 0)
  $\mathbf{then}$
   selectedDrink := drink
  $\mathbf{end}$
  
  $\mathbf{vend\_soda}$ $\mathbf{refines}$ $\mathbf{vend}$ $\triangleq$  $\mathbf{when}$
   selectedDrink = soda
  $\mathbf{then}$
   soda :=  soda - 1 ||
   selectedDrink := none ||
   coin := coin - 1
  $\mathbf{end}$

  $\mathbf{vend\_water}$ $\mathbf{refines}$ $\mathbf{vend}$ $\triangleq$ $\mathbf{when}$
    selectedDrink = water
  $\mathbf{then}$
    water :=  water - 1 ||
    selectedDrink := none ||
    coin := coin - 1
  $\mathbf{end}$

  $\mathbf{restock}$ $\mathbf{refines}$ $\mathbf{restock}$ $\triangleq$  $\mathbf{when}$
    soda + water = 0
  $\mathbf{then}$
    soda :=  2 ||
    water :=  1 
  $\mathbf{end}$
\end{lstlisting}
\end{minipage}
\end{figure}

\subsubsection{Formalization}
\label{subsec:formalization}
A B machine induces a state space, as formalized in \Cref{def:machine}, similar to Zhu et al.~\cite{Zhu2019}.
\begin{definition}
\label{def:machine} (State Space of a machine)
    The state space of an Event-B machine M is a quadruple (S, E, $\delta$, $I$) where
    S is the 
    set of all possible states, $I \subseteq S$ the initial states,
    E is the set of events, 
    and $\delta \in E \tfun (S \rel S)$ describes the transitions. 
    To denote an individual transition, we write $s \overset{e}{\rightarrow}_{M} s' $ if $(s \mapsto s') \in \delta(e)$.
\end{definition}

A state itself consists of a mapping from variables and constants to values.

Note that later some propositions will require finite state spaces.\footnote{\Cref{def:prop:convergence_induces_stable_states,def:prop:mainProp} will rely on this.}




\begin{example*}
    For our machine in \Cref{lst:vending}, the state space is the
    quadruple (S, E, $\delta$, I) with
    S = $\{ \{soda = 3, coin = 0 \} , \{soda = 3, coin = 1 \},...\}$,\\
    I = $\{ \{soda = 3, coin = 0 \} \}$,
    E = $\{ insert\_coin, vend, restock \}$, and \\
    $\delta = \{ insert\_coin \mapsto \{(\{soda = 3, c = 0 \}, \{soda = 3, coin = 1 \}),... \},... \} $. \\
    We have, e.g., 
    $\{soda = 3, coin = 0 \} \overset{insert\_coin}{\rightarrow}_{M} \{soda = 3, coin = 1 \}$.
\end{example*}

\begin{definition} (Event-B refinement)
\label{def:ref}
    Let ($S_A$, $E_A$, $\delta_A$, $I_A$\footnote{For the following we assume $|I_A|=1$. }) and ($S_C$, $E_C$, $\delta_C$, $I_C$) be the state space of two machines $A$ and $C$.
    We write A $\sqsubseteq_{ref}$ C if C is an Event-B refinement \cite{Abrial2010} of A.
    During refinement, new events can be introduced in $C$, which refines the invisible \texttt{skip}. 
    We denote the set of these events as $\mathcal{N}_C$. 
    If no new events are introduced, $\mathcal{N}_C = \emptyset$.
    Event-B also allows renaming events.
 We capture this renaming in the
     mapping  $\psi_{C,A} \in E_C \setminus \mathcal{N_C} \tsur E_A$ from concrete to abstract events.%
\footnote{$\tsur$ is the total surjection.}
    This means that every event, minus the events introduced via \texttt{skip}, is assigned to an abstract event.
\end{definition}

Note that the definition of $\psi_{C,A}$ as a function above has subtle consequences.
If an abstract event has parameters that do not exist at the concrete level, Event-B stipulates the definition of a witness predicate (see Section 5.1.7 of \cite{Abrial2010}).
To make $\psi_{C,A}$ unambiguous (i.e., a function), we require that the witness predicate
 is {\em deterministic}, i.e., it has a single solution for every concrete event.
This requirement will play an important role later in our main Proposition~\ref{def:prop:mainProp}.
In practice, most witnesses are deterministic, so our requirement is reasonable.

    We can use the relational image operator [~]  to apply $\psi_{C,A}$ to sets, i.e, $ \psi_{C,A}[E] = \{e_c\in E \mid \psi_{C,A}(e_c)\}$.

\subsubsection{Traces in Event-B}
\label{subsec:tracesEventB}
Another view on Event-B machines is from the perspective of traces, which consist of states and events.
\Cref{eq:trace} is aligned with the definition of Zhu et al.~\cite{Zhu2019}.  

\begin{definition}
\label{eq:trace} (Finite trace)
Let the state space of a machine M be ($S_M$, $E_M$, $\delta_M$, $I_M$).
We define the transitive closure of  $s \overset{e}{\rightarrow} s' $  as follows:
$s_0 \overset{e_0, e_1,...,e_n}{\implies} s_n$ if $s_0 \overset{e_0}{\rightarrow} s_1 \overset{e_1}{\rightarrow} ..\overset{e_n}{\rightarrow} s_n $. 
Let $E^{*}_{M}$ be the set of finite sequences over $E_M$.
We denote the set of all traces of M as follows:
$traces(M) = \{\sigma \in  E^{*}_{M} \mid s_0\in I_M \wedge s_n \in S_M\}$.
Finally, we denote the last state reached by a trace $\sigma$ by $last(\sigma)$.
\end{definition}

\begin{example*}
   Consider \Cref{fig:trace1}, where we see a visualization of the beginning of the state space of \Cref{lst:vending}.
   This part of the state space has the following traces: 
$\langle  \rangle $, $\langle  insert\_coin \rangle, \langle insert\_coin, insert\_coin \rangle,  \langle insert\_coin, vend \rangle $, etc. 
Note that we do not explicitly write the \texttt{INITIALISATION} event, rather assuming it has been performed implicitly at the beginning of a trace to reach the initial state. 
\end{example*}


\section{Traces and failure traces in Event-B}
\label{sec:traces_failure}
\subsection{Trace refinement}

Trace refinement is usually defined via the subset relation over the set of traces \cite{Roscoe1998} ($A \sqsubseteq_t C$ iff $traces(C) \subseteq traces(A)$).
However, due to the renaming $\psi_{C,A}$ and new events $\mathcal{N}_C$, we cannot apply this definition as is.
In the following Definition \ref{def:renaming}, we take these particularities into account to be able to compare the traces of $A$ and $C$:
\begin{definition} (Concealment \& renaming in traces)
\label{def:renaming}
Let two machines, A and C, be in a refinement relationship $A \sqsubseteq_{t} C$. 
For $\sigma = e_1,\ldots,e_n \in traces(C)$ we define:
\begin{enumerate}
\item $\tau_{C,A}(\sigma)= \langle \rangle$ if $n=0$
\item $\tau_{C,A}(\sigma) = \tau_{C,A}(e_2,\ldots,e_n)$ if $e_1 \in \mathcal{N}_C$
\item $\tau_{C,A}(\sigma) = \psi_{C,A}(e_1) \frown \tau_{C,A}(e_2,\ldots,e_n)$ otherwise.
\end{enumerate}
The $\tau_{C,A}$ operator removes the skip events from the trace and renames the remaining events to their counterparts in $A$.

$\tau$ is also applicable to a set of traces as follows $\tau_{C,A}(T) = \{ t \in T | \tau_{C,A}(t)\}$.
\end{definition}

With the help of $\tau$ we can give a notion of trace refinement based on our definition of trace \Cref{eq:trace}.
\begin{definition} (Trace refinement)
\label{def:trace_ref} 
A is a trace refinement of C, denoted by A $\sqsubseteq_{t}$ C,  iff $ \tau_{C,A}(traces(C)) \subseteq traces(A)$.
\end{definition}

\begin{figure}
    \centering
    \includegraphics[width=0.65\textwidth]{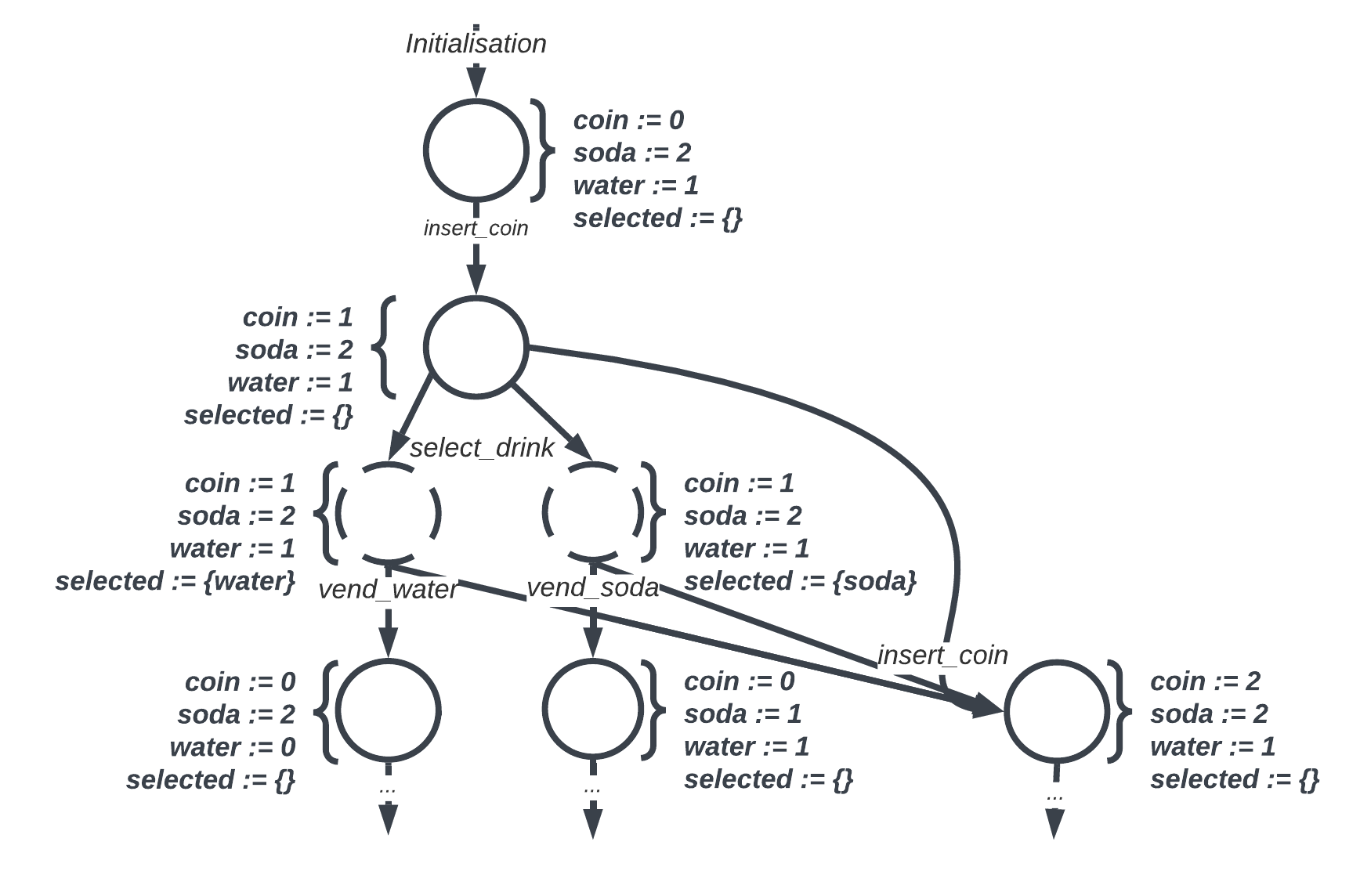}
    \caption{Visualization of the state space of \Cref{lst:vending2}}
    \label{fig:trace2}
\end{figure}

\begin{example*}
Consider \Cref{fig:trace2}, which shows a part of the state space of \Cref{lst:vending2}.
The traces of this machine would be the following: 
$ \langle \rangle, \langle insert\_coin \rangle, $\allowbreak$\langle insert\_coin, select\_drink \rangle$, etc. Concealing and renaming yields: \\
$\tau_{m1,m0}(\langle  insert\_coin, select\_drink \rangle) = \langle  insert\_coin \rangle$.
\end{example*}

Finally, we know from Derrick and Boiten~\cite{Derrick2018} that trace refinement is strong enough to mimic forward-backward simulation (it is more strict than pure forward simulation and, consequently, stricter than standard Event-B POs). This insight has multiple consequences for us: 1) A refinement relationship built on top of trace refinement will be, at minimum, as strong as forward-backward simulation, and 2) all Event-B POs will be preserved. 
Every concrete trace thus has an abstract equivalent in trace refinement.
However, in our general aim to preserve liveness properties, it also becomes clear that we need a more powerful refinement notion to show that every abstract trace has a concrete equivalent.

\subsection{Failure traces in Event-B}
\label{sec:failure_refinement}
Failure traces~\cite{Derrick2018,Hoare1985} extend regular traces by recording additional information about their last state, namely which events are refused at the end. 

\begin{definition} (Enabled \& refused events)
\label{eq:definition_enabled}
Let $s \in S_M$ be a state of M. The enabled events of $s$ are defined by 
$enabled(s) = \{ e \mid (s \mapsto s') \in \delta_M(e) \} $.
The refused events are the complement: $\mathit{refusal}(s) = E_M \setminus \text{enabled}(s)$.   
\end{definition}

Now, we additionally need a notion of stable states to only compare the proper states in a refinement check, i.e., those concrete states that do not allow the execution of events introduced by \texttt{skip}
\begin{definition} (Stable states)
\label{eq:definition_stable}
Let $A \sqsubseteq_{t} C$.
We define $stable_C = \{ s \in S_C \mid \nexists e \cdot e \in \mathcal{N}_C \land ( s \mapsto s') \in \delta_C(e)\}$,
 where $S_C$ are the states of $C$.
This means a state has no outgoing transitions introduced by refining \texttt{skip}.
A state s is stable iff $s \in stable_C$.
If C is not a refining machine, then all states of C would be stable.
\end{definition}

We use the notion of traces, refused events, and stable states to create the notion of failure trace in \Cref{eq:definition_failure}.
\begin{definition} 
\label{eq:definition_failure} (Failure trace) The set of failure traces for a machine M are defined as $failure(M) = \{(\sigma,X) \mid \sigma \in traces(M) \land last(\sigma) \in stable_M \land X = \mathit{refusal}(last(\sigma))\}$
\end{definition}

\begin{example*}
\label{exmp:failure1}
Again, we consider \Cref{fig:trace1} and \Cref{lst:vending}. From the visible state space, we have the following failure traces: $(\langle \rangle, \{vend, restock \}), (\langle insert\_coin \rangle,$\allowbreak$ \{  restock \}) $, etc. 
\end{example*}

\subsection{Failure divergence refinement}
\Cref{eq:definition_failure} is stronger than \Cref{eq:trace} -- we store trace plus additional information.
The next goal is to define refinement for failure traces.
For this, we must translate failure sets while accounting for Event-B particularities. We introduce a particular translation operator in \Cref{def:absRefusals}.
Note that this function not only takes a set of events and renames them but also ensures that all refining events are present for each refined event.
This is necessary so as not to wrongly discard intuitively correct failure refinements. We give an example for this case with the notion of \Cref{def:failure_trace_ref}.

\begin{definition}
    \label{def:absRefusals}
Let $X$ be a refusal set within the context of two machines $A \sqsubseteq_{t} C$. The set of abstract refusals is defined as:
   $\mathit{AbsRefusal}(X) = \{y \in \psi_{C,A}[X] \mid \psi_{C,A}^{-1}[{y}] \subseteq X\}$. Furthermore, as the last state is stable $X \cap \mathcal{N}_C = \emptyset$.
\end{definition}

With this, we define the failure trace refinement in \Cref{def:failure_trace_ref}.
\begin{definition} (Failure trace refinement)
\label{def:failure_trace_ref} We define failure trace refinement as $\forall (\sigma, X)  \cdot (\sigma, X) \in failure(C) \limp (\tau_{C,A}(\sigma), \mathit{AbsRefusal}(X)) \in failure(A)$ and write $A \sqsubseteq_{f} C$.
\end{definition} 

\begin{example*}
Consider \Cref{fig:trace2}, which shows a part of the state space of \Cref{lst:vending2}. The failure traces of this machine would be the following
$ (\langle \rangle, $\\ $ \{ vend\_water, vend\_soda, restock \}) $ with $\mathit{AbsRefusal} $ being $\{ vend, restock \}$, \newline
$ (\langle insert\_coin, select\_drink \rangle,\{restock, vend\_soda\}) $ with $\mathit{AbsRefusal} $  $ \{ restock \}$.
Applying $\tau$ to the trace part yields the traces from \Cref{{exmp:failure1}}. \newline
$(\langle  insert\_coin \rangle, \{vend, restock \})$ is not a valid failure trace, as the last state is unstable. 
Elaborating on the necessity of stable states for failure traces, we consider the following example: In \Cref{fig:trace2},  a trace $\pi = \langle insert\_coin \rangle$ would have the $\mathit{refusal}(last(\pi))$ = \{vend\_water, vend\_soda, restock\}. However, $\mathit{refusal}(last(\tau(\pi))) = \{restock\}$ misses the $vend$ transition. Thus by only using this weak definition, \Cref{lst:vending2} would not refine \Cref{lst:vending}.

Motivating \Cref{eq:definition_stable} consider $\sigma = \langle insert\_coin \rangle$. If we do not force stable states, \Cref{lst:vending2} would not refine \Cref{lst:vending}, as the failure sets are different. 

For \Cref{def:absRefusals}, consider $\sigma $=$ \langle insert\_coin, select\_drink \rangle$ with $\mathit{refusal}(last(\sigma))$ = $\{$restock, vend\_water$\}$. Consequently, $\tau(\sigma) = \langle insert\_coin \rangle$ with $\mathit{refusal}(last(\tau(\sigma)))$ = $\{$restock$\}$. Both traces would refine each other but not failure refine each other, even though they represent the same (abstract) behavior.

\end{example*}
Event-B can introduce divergence during refinement by presenting events that refine skip. Usually, divergence is prevented by marking events with the keywords \texttt{convergent} or \texttt{anticipated}, which induce POs that must be discharged to establish that these events do not introduce divergence. We check for this property without relying on these keywords, as shown in \Cref{eq:definition_convergent}.

\begin{definition} (Divergence and failure divergences refinement)
\label{eq:definition_convergent}
Let $A \sqsubseteq_{t} C$ and $C$ have a finite state space.
We define the set of divergent states of $C$ by
$divergence_C =  \{ s \mid \exists \sigma \in \mathcal{N}_C^{*}$\footnote{Similar to $E_M^{*}$ back in \Cref{eq:trace}.} . $s \overset{\sigma}{\implies} s \} $.
I.e., such states from which we find a sequence of \textit{skip} refining events such that they form a loop.
A trace refinement free of divergence is a trace divergence refinement $A \sqsubseteq_{td} C$ iff $A \sqsubseteq_{t} C \land  divergence_C = \emptyset$.
A failure trace refinement free of divergence is a failure divergence refinement $A \sqsubseteq_{fd} C$ iff  $A \sqsubseteq_{f} C \land divergence_C = \emptyset$.
\end{definition}

It could happen that a refining machine only consists of unstable states, and thus, we could not make statements about failure (divergence) refinement. To rule out this phenomenon, we introduce a guarantee for stable states in machines.
\begin{proposition}[Divergence freeness guarantees the existence of stable states]
\label{def:prop:convergence_induces_stable_states}
Let $A \sqsubseteq_{td} C$ and the state spaces be finite.
Further, let $\sigma \in traces(A)$ and 
let $\Pi = \{\pi \in traces(C) \mid \tau_{C,A}(\pi) = \sigma \}$.
If $\Pi \neq \emptyset$ then $\exists \pi  \cdot \pi \in \Pi$ such that $last(\pi) \in stable_C$. 
\end{proposition}

With these tools, we can now show that, under some conditions, abstract traces have a concrete counterpart under failure divergence refinement.

\section{Trace equivalence}
\label{sec:formalization}

\newcommand{\intchoice}{\;\sqcap\;}
\newcommand{\extchoice}{\;\Box\;}
As a failure trace also records disabled events, failure refinement
 cannot simply remove choices that are present in the abstract machine.
Another observation is that failure divergence refinement in Event-B works differently from the CSP version. 
In CSP, failure divergence refinement does not imply trace equivalence:
\begin{example*}
Let C = $a \rightarrow STOP$ and A = $a \rightarrow STOP \intchoice b \rightarrow STOP$ be two CSP processes.%
\footnote{
Observe that $\intchoice$ is the internal choice operator that induces an invisible skip (aka $\tau$) action in CSP.}
We have the following failure traces
\begin{enumerate}
     \item $failures(C) = \{ (\langle\rangle,\{b\}), (\langle a\rangle,\{a,b\}) \}$,
     \item $failures(A) = \{ (\langle\rangle,\{b\}), (\langle a\rangle,\{a,b\})  \} \cup 
      \{ (\langle\rangle,\{a\}), (\langle b\rangle,\{a,b\}) \} $.
\end{enumerate}
Hence $C$ failure refines $A$, i.e., $A \sqsubseteq_{fd} C$, but the trace $\langle b \rangle \in traces(A)$ has no counterpart in $C$.
So, in this case, failure refinement does not ensure the preservation of traces.
This counter-example does not carry over to our Event-B setting.
Indeed, a major difference is that in our setting, skip events can only occur in the refined model, not
 in the abstract one. 
\end{example*}

Unfortunately, the above example using the internal choice operator in the abstraction is not the
 only problem, as the following example shows.
\begin{example*}
Let C = $a \rightarrow STOP$ and 
    A = $a \rightarrow b \rightarrow STOP \extchoice a \rightarrow STOP$ be two CSP processes
    (where $\extchoice$ is the external choice operator).

\begin{enumerate}
     \item $failures(C) = \{ (\langle\rangle,\{b\}), (\langle a\rangle,\{a,b\}) \}$,
     \item $failures(A) = \{ (\langle\rangle,\{b\}), (\langle a\rangle,\{a,b\})  \} \cup 
      \{ (\langle a\rangle,\{a\}), (\langle a,b\rangle,\{a,b\}) \} $.
\end{enumerate}

Again $C$ failure refines $A$, i.e., $A \sqsubseteq_{fd} C$, but the trace $\langle a,b \rangle \in traces(A)$ has no counterpart in $C$.
This time, though, the example does translate to Event-B. Listing~\ref{lst:nondet} shows the Event-B encoding of A, and Listing~\ref{lst:concrete} shows C.
The fundamental problem is that $A$ is not a deterministic event (see Definition~\ref{def:evdet} below). Indeed, in our proof later, we need to map a failure trace in $C$ to a single failure trace in $A$. Our solution is to disallow machines like the one in Listing~\ref{lst:nondet}.
Luckily, every Event-B machine can be rewritten into an event-deterministic one, possibly by adding parameters.
This is done in Listing~\ref{lst:evdet}.
Note that if the concrete machine does not also have this parameter, Event-B requires the addition of a witness predicate. Here, we also require that the witness predicate produces a single solution.%
\footnote{This is usually the practice case; witness predicates are always equalities.}
%
\end{example*}

\begin{figure}
\noindent\begin{minipage}{.30\textwidth}
\begin{lstlisting}[mathescape, label={lst:nondet}, caption={Non-deterministic abstract machine}, captionpos=b, numbers=left, stepnumber=1, frame=single,framexleftmargin=2.2em, framexrightmargin=-1.5em]
$\mathbf{machine}$ A 
$\mathbf{variables}$ 
  pc $\in$ 0..3
$\mathbf{events}$
  INITIALISATION $\mathbf{then}$
   pc := 0
  $\mathbf{end}$

  $\mathbf{a}$ $\triangleq$ $\mathbf{when}$
   pc=0
  $\mathbf{then}$
   pc :: {1,2}
  $\mathbf{end}$

  $\mathbf{b}$ $\triangleq$ $\mathbf{when}$
    pc=2
  $\mathbf{then}$
    pc := 3 
  $\mathbf{end}$
  
\end{lstlisting}
\end{minipage}\hfill %
\noindent\begin{minipage}{.30\textwidth}
\begin{lstlisting}[mathescape, label={lst:concrete}, caption={Concrete refinement}, captionpos=b, numbers=left, stepnumber=1, frame=single,framexleftmargin=2.2em, framexrightmargin=-1.5em]
$\mathbf{machine}$ C $\mathbf{refines}$ A
$\mathbf{variables}$ 
  pc $\in$ 0..3
$\mathbf{events}$
  INITIALISATION $\mathbf{then}$
   pc := 0
  $\mathbf{end}$

  $\mathbf{a}$ $\triangleq$ $\mathbf{when}$
   pc=0
  $\mathbf{then}$
   pc := 1
  $\mathbf{end}$

  $\mathbf{b}$ $\triangleq$ $\mathbf{when}$
    pc=2
  $\mathbf{then}$
    pc := 3 
  $\mathbf{end}$
  
\end{lstlisting}
\end{minipage}\hfill %
\begin{minipage}{.30\textwidth}
\begin{lstlisting}[mathescape, label={lst:evdet}, caption={An event-deterministic machine}, captionpos=b, numbers=left, stepnumber=1, frame=single,framexleftmargin=2.2em, framexrightmargin=-1.5em]
$\mathbf{machine}$ ADet 
$\mathbf{variables}$ 
  pc $\in$ 0..3
$\mathbf{events}$
  INITIALISATION $\mathbf{then}$
   pc := 0
  $\mathbf{end}$

  $\mathbf{a}$ $\triangleq$  $\mathbf{any}$ npc $\mathbf{where}$
   pc=0  $\land$ 
   npc $\in$ {1,2}
  $\mathbf{then}$
   pc := npc
  $\mathbf{end}$

  $\mathbf{b}$ $\triangleq$  $\mathbf{when}$
    pc=2
  $\mathbf{then}$
    pc := 3 
  $\mathbf{end}$
\end{lstlisting}
\end{minipage}
\end{figure}

The above insight leads to the following definition:

\begin{definition}  \label{def:evdet}
An Event-B machine $M$ is {\bf event deterministic} if $\forall \sigma, X_1,X_2$ we have
  $\{(\sigma,X_1),(\sigma,X_2)\} \subseteq failure(M) \Rightarrow X_1=X_2$.
\end{definition} 
This is guaranteed in Event-B if we only use deterministic assignments.
Note that one can always rewrite an Event-B model into such a form.
E.g., for $x::E$ we add a parameter $x'$ with $x'\in E$ in the guard and use the assignment $x := x'$.
This was done in Listing~\ref{lst:evdet}.

As discussed in Definition~\ref{def:ref} above, we require $\psi_{C,A}$ to consider the witness predicates, and we require the witness predicates to be deterministic. Together with Definition~\ref{def:evdet}, we can thus map any concrete trace to a single abstract trace with a single failure set, which allows us to prove our main result in Proposition~\ref{def:prop:mainProp}.

To extend the power of our primary contribution, we extend our definition of traces one last time to infinite traces in \Cref{def:infi_traces}.
\begin{definition} (Infinite traces)
\label{def:infi_traces}
    Let the state space of a machine M be ($S_M$, $E_M$, $\delta_M$, $I_M$).
    Let $E_M^{\Delta}$ the set of infinite sequences, i.e., $s_0 \overset{e_0, e_1,...}{\implies}$.
    We denote the infinite traces starting from an initial state as $traces(M)_\Delta$. 
\end{definition} 
With the notion of failure divergence refinement, traces, and infinite traces, we can now reason whether a refinement preserves finite and infinite traces.

\begin{proposition} [Failure divergence refinement implies trace equivalence]
\label{def:prop:mainProp}
    Let A and C be two machines with finite state spaces and with $A \sqsubseteq_{fd} C$. 
    Let A be event deterministic.
    Then $traces(A) \subseteq \tau_{C,A}(traces(C))$, and $traces(A)_\Delta \subseteq \tau_{C,A}(traces(C)_\Delta)$\footnote{$\tau_{C,A}$ is the renaming and concealment operator from \Cref{def:renaming}.}. I.e., failure divergence refinement guarantees that every abstract trace has a concrete equivalent.
\end{proposition}

This proposition implies that all arguments and reasoning for trace properties in the abstract machine carry over to later refinements\footnote{The backward direction of the proven inclusion is enforced by trace refinement/forward-backward simulation.}. This simplifies the validation process, allowing us to validate trace properties in their abstract forms, facilitating early validation, and reducing development overhead.


\section{Failure divergence checking in practice}
\label{sec:approach}
We build on existing tools to realize failure divergence refinement for Event-B. ProB~\cite{Leuschel2003} is an animator and model checker for formal modeling languages like Event-B. 
We will use ProB as a base for our algorithm as it can understand and interact with Event-B models, helping us focus more on implementing the approach than handling the models.

Before we implemented our approach, ProB could already deal with trace refinement, as shown in the previously cited contribution by Leuschel and Butler~\cite{Leuschel2005}. For this, the user must first model-check the abstract model, export the state space, and import the abstract state space and concrete machine into the algorithm for trace refinement check. Our work will use this existing infrastructure as it is reliable and robust enough to handle the extension to Event-B failure divergence refinement.

\subsection{Algorithm}
\label{subsec:algorithm}
\paragraph{Setup.}
The failure divergence algorithm is shown in \Cref{alg:algB}. The algorithm's input is the $\psi_{C,A}$, 
and $\mathcal{N}_{C}$, 
which we must synthesize beforehand. Furthermore, we expect both machines' state space to be available to access them with the $trans(s_1, e,s_2)$ predicate. This predicate works via unification\footnote{Like the Prolog programming language.}, e.g., we provide it with two states, and it will infer the possible events that satisfy the predicate. 

For the output, we expect the evaluation's result to be either true or false. If both machines are not valid failure divergence refinements, we expect an abstract and concrete trace that serves as a counterexample.

\paragraph{Computation.}
The core idea is to uncover all reachable pairs of abstract and concrete states (\cref{alg:iterateOverConcreteEvents,alg:abstConcreteComparision}) while checking for every pair of states that the failures are the same (\cref{alg:failure,alg:failureRef}). The function in \cref{alg:failureRef} also checks for divergence\footnote{We abstracted the simpler divergence check for the sake of readability and space usage. In practice, the divergence check is part of the algorithm.}. \Cref{alg:stableCheck} checks whether the current state is unstable, and if this is the case, we skip the computation of the abstract trace. The algorithm recursively iterates over all reachable transitions (\cref{alg:loop1,alg:loop2}) and stores the encountered pairs (\cref{alg:addSeen}), thus achieving complete coverage. Should the algorithm find that a pair of states is not a failure refinement of each other, it will terminate with a \texttt{false}. If the algorithm can visit all reachable states in both machines without finding any nonmatching pairs, the algorithm terminates with a \texttt{true}.

\paragraph{Limitations.}
The algorithm works on finite state spaces, and as the set of seen states increases in every recursion, eventually, the algorithm terminates. Practically, the modeler is limited by available memory as all possible computable transitions are used. 

The problem of failure/divergence is generally seen as a PSPACE-hard problem (see Appendix C.1 in Roscoe~\cite{Roscoe1998}). 
Currently supported are only deterministic choices in the initial state, i.e., cases where $|I_a| = 1$ and $ |I_c| = 1$. Furthermore, the algorithm expects both machines to have some refinement relationship as it has to calculate the $\psi_{C,A}$ function and the $\mathcal{N}_{C}$ set. 

Another limitation is that we have set $\delta$ as a total function back in \Cref{def:machine}. This means we do not allow multiple transitions between two states. Finally, we currently restrict ourselves to refinement that uses the \textit{extends} keyword for our implementation. For full support for the \textit{refines} keyword, we lack support for reordering, redefinition, renaming, or removing parameters. However, it is a technical problem to draw this information out of the AST of the Event-B machine and preprocess it to make it available.
\setlength{\textfloatsep}{0pt}
\begin{algorithm}[t]
\scriptsize
\DontPrintSemicolon
\KwIn{ The state space of the abstract machine accessible via $trans_a(s_1$, $e$, $s_2$) and the state space of the concrete machine via $trans_c(s_1$, $e$, $s_2$), furthermore the relation $\psi_{C,A}$ and a set of $\mathcal{N}_{C}$ events. Finally, we have a set of seen states $S_s$ that is empty at the start}

isFailureRefinement $:=$ \text{failLoop}($s_c$, $s_a$, $\langle \rangle$, $\langle \rangle$) \;
 
\SetKwBlock{Begin}{function}{end function}

\Begin($\text{failLoop} {(}s_c, s_a, t_c, t_a{)}$){ \label{alg:primaryF}
\uIf{$(s_c, s_a) \notin S_s$}{ \label{alg:visitied} 
    $S_s := S_s \cup \{(s_c, s_a)\}$\; \label{alg:addSeen}
    \uIf{failureRefines($s_c$, $s_a$)}{ \label{alg:failure}
        nexCon $:= \{ (e_c, s_c') \mid trans_a(s_c, e_c, s_c') \} $ \; \label{alg:nexCon}
        \ForAll{$(e_c, s_c') \in nexCon$  }{ \label{alg:iterateOverConcreteEvents}
            \uIf{$e_c \in \mathcal{N}_{C}$}{ \label{alg:stableCheck}
                $t_c^{tp} := t_c +(e_c, s_c') $ \;
                res $:=$ \text{failLoop}($s_c'$, $s_a$, $t_c^{tp}$, $t_a$) \; \label{alg:loop1}
                \If{$(false, t_c', t_a') = res $}{
                    \Return{ $ (false, t_c', t_a')$}
                }
            }
            \Else{
                nexAbs $:= \{ (e_a, s_a') \mid trans_a(s_a, e_a, s_a') \}$\; \label{alg:failureSetAbst}
                res $:= \emptyset$ \;
                \ForAll{$(e_a, s_a') \in nexAbs \land e_a = \psi_{C,A} (e_c)$ }{  \label{alg:abstConcreteComparision} 
                    $t_c^{tp} :=  t_c+ (e_c, s_c') $ \;
                    $t_a^{tp} := t_a + (e_a, s_a')  $ \;
                    res $:=$ res $\cup$ \text{failLoop}($s_c'$, $s_a'$, $t_c^{tp}$, $t_a^{tp}$)\; \label{alg:loop2}
                }
                \If{$\forall (b, t_c', t_a') \cdot (b, t_c', t_a') \in res \land b = false$ }{
                    \Return{any $ (false, t_c', t_a') \in res$}
                }
            }
        }
        \Return{ $ (true, t_c, t_a) $} \label{alg:sideResultTrue}
    }
    \Else{
        \Return{$(false, t_c, t_a) $} \label{alg:NotFailure}
    }
}
\Else{
 \Return{$(true, t_c, t_a)$ } \label{alg:finalResultTrue}
}
}
\Begin($\text{failureRefines} {(}s_c, s_a {)}$){ \label{alg:failureRef}
    \uIf{$\nexists e  \cdot e \in \mathcal{N}_{C}  \land trans_c(s_c, e, \_s_c')$}{ \label{alg:forallEventsInFailure}
        absFailure  $:= \{e_a \mid trans_a(s_a, e_a, \_s_a') \}$ \;
        conFailure  $:= \{e_c  \mid trans_c(s_c, e_c, \_s_c') \}$ \;
        \Return{absFailure = AbsRefusals(conFailure)} 
    }\Else{
        \Return{ $s_c \in divergence_C$ } \label{alg:convergesCheck}
    }    
}
\caption{Failure divergence refinement checking}
\label{alg:algB}

\end{algorithm}

\section{Evaluation}
\label{sec:evaluation}
For evaluating our algorithm, we have two aims\footnote{A replication package can be found at:\\ \url{https://zenodo.org/doi/10.5281/zenodo.10377564}. It contains a stand-alone version of the tool and the files used in the case study.}. First, we want to show that the algorithm works and can provide an advantage by establishing a failure divergence refinement relationship or pointing out mistakes. Second, we want to do this within a reasonable time compared to other similar operating validation methods. We consider methods similar if they also explicitly traverse the entire state space. This paper used basic model checking, i.e., checking for invariant violations and trace refinement checking, a contribution of Leuschel and Butler~\cite{Leuschel2005}. We extended the provided algorithm to handle Event-B particularities as discussed in \Cref{subsec:eventb_ref}. Model checking and trace refinement are provided within ProB\footnote{FDR3, as the prominent refinement checker for CSP does not support Event-B; therefore, we cannot perform a direct comparison.}. 

As a model to evaluate, we choose our example model from \Cref{lst:vending,lst:vending2} a version of the interlocking Event-B model introduced by Stock et al.~\cite{Stock2023} and a version of the hemodialysis machine (HD machine) Event-B model introduced by Hoang et al.~\cite{Hoang2016a}.

\subsection{Case studies}
\paragraph{Interlocking}
Interlocking describes a system managing trains and train tracks in a rail yard. Routes through the rail yard are divided into blocks they share with other routes. The peculiarity is that trains should not collide and should be on a valid track configuration. The model here was initially introduced by Abrial~\cite{Abrial2010}, and the authors in Stock et al.~\cite{Stock2023} extend it to demonstrate the so-called abstraction approach. 


\paragraph{Hemodialysis machine}
Hemodialysis purifies a patient's blood if the patient's body cannot do it by itself, e.g., in case of kidney failure. For the hemodialysis process, a machine controls parameters such as heart rate and blood pressure and adjusts the treatment accordingly. The case study was proposed by Mashkoor~\cite{Mashkoor2016}, and we use the implementation of Hoang et al.~\cite{Hoang2016a}.

\subsection{Experiment setup}
We ran all models through the ProB~\cite{Leuschel2003} model checker. We used an Intel Core i7-10700 2.90GHz × 8 CPU with 16GB RAM running Linux Mint for the benchmarks. Each measurement was run ten times, and the mean was taken. We restricted the search space to make it finite for our evaluation of the HD machine. The model-checking results are shown in \Cref{tab:baseline}. We measured the time spent and the number of states and transitions to give insight into the model's size. 

Seven machines were investigated, the two vending machines from the beginning (vend\_abst \& vend\_conc) with abstract and concrete machines, respectively. There are three versions of the interlocking (inter\_abst \& inter\_conc \& inter\_fixed), with inter\_fixed being a corrected version of inter\_abst. We found a bug during failure divergence refinement checking and provided a fix. Finally, we selected sizable models from the HD machines (hd\_abst \& hd\_conc), again with the first being the abstract and the second being the refinement. 


\begin{table}
\centering
\begin{tabular}{l|rrr}
Project &
  \begin{tabular}[c]{@{}r@{}}Model checking\\ \end{tabular} &
  \begin{tabular}[c]{@{}r@{}}Number \\of States\end{tabular} &
  \begin{tabular}[c]{@{}r@{}}Number of \\ Transitions\end{tabular} \\ \hline
   vend\_abst    & 26 ms    & 10    & 14      \\
   vend\_conc  & 38 ms    & 29    & 47      \\ \hline
   inter\_abst   & 41 ms    & 81    & 325     \\
   inter\_fixed  & 33 ms    & 56    & 186     \\
   inter\_conc     & 3666 ms  & 3508  & 11714   \\ \hline
    hd\_abst         & 484 ms & 484 & 3282     \\
   hd\_conc            & 11218 ms & 16968  & 94738  \\
\end{tabular}
\caption{Baseline performance of model checking on the models}
\label{tab:baseline}
\end{table}

\subsection{Results}
\paragraph{Vending machine.}
For the vending machine model, we could successfully establish failure divergence refinement. The runtime is short, as shown in \Cref{tab:comparision}. The individual runs had high fluctuations, and one reason for this may be the small size and internal optimization strategies. Therefore, we could not see a clear trend in which technique is faster or slower. 

\paragraph{Interlocking.}
Initially, the interlocking model failed the failure divergence refinement check (see \Cref{tab:comparision}). Using counterexamples provided by our tool, we identified a divergence issue caused by an event refining \texttt{skip} while being enabled infinitely without altering the state. As we saw no good reason for this behavior, we prohibited it by strengthening the event guards.

Additionally, we observed that the refinement was stricter concerning allowed route combinations, concluding that block abstraction was too liberal. To rectify this, we reintroduced a restriction on route exclusivity without relying on blocks, successfully validating our fixed model.

The latter fix shrinks the state space significantly, as several route combinations are now invalid. This can be seen in \Cref{tab:comparision}. Compared to \Cref{tab:baseline} the refinement checking is faster than model checking for the fixed version.

\paragraph{Hemodialysis machine.}
Learning from the interlocking, we did a brief pre-check for divergence for the HD machine.
One event to set the patient's values 
introduced divergence by being enabled constantly. We altered the model so that the event could only be used once. With this modified model, we established failure divergence refinement. The statistics can be seen in \Cref{tab:comparision}.  

The trace refinement is significantly faster than failure divergence refinement. We explain this by the fact that the model has the following three new events: \texttt{PM\_SetsPressure}, \texttt{CS\_LowLevel\_Abnormal}, and \texttt{CS\_TopLevel\_RaisesAlarm}. For those events, divergence must be checked in nearly every state, as almost every state is stable. The divergence check is expensive as it uses set operations internally. This might be a direction for future improvement.

\paragraph{General observations.}
Sometimes, model checking is slower than trace and failure divergence refinement checking. The internal mechanism of ProB and the nature of the models explain this. Calculating a specific transition can be more expensive than other transitions, depending on the complexity of the constraints describing the transitions. Similarly, during refinement checking, the exploration process is guided by the already explored abstract state space, which can reduce the calculation cost.

\begin{table}
\centering
\begin{tabular}{l|rr}
Project & \begin{tabular}[c]{@{}r@{}}Trace refinement\\ \end{tabular} & \begin{tabular}[c]{@{}r@{}}Failure divergence \\refinement \end{tabular} \\ \hline
Vending Machine & 51 ms     & 46  ms  \\
Interlocking    & 4077 ms   & 19 ms\tablefootnote{Results in a fail (divergence).}  \\
Interlocking (fixed)    & 934 ms   & 1008 ms  \\
HD machine           & 560 ms  & 4147 ms
\end{tabular}
\caption{Comparison trace refinement and failure divergence refinement}
\label{tab:comparision}
\end{table}

\subsection{Lessons learned}
From our experiments, we could learn multiple things. \textbf{First}, failure divergence refinement is often slower than trace refinement checking as divergence and failure sets need to be calculated. The divergence and refusal sets calculations can be expensive, as we can see with the HD machine.
\textbf{Second}, counterexamples provided helpful information in cases where failure divergence refinement failed. We could spot subtle undetected issues, e.g., divergence, which many case study implementations do not explicitly treat. 

\subsection{Threats to validity}
\label{subsec:threats}
Based on the categories provided by Wright et al.~\cite{Wright2010}. We want to highlight a threat to internal validity. As we relied on ProB, we only used those models we knew ProB could handle in the first place (i.e., those that can be fully explored). Furthermore, we only choose those models that have a deterministic initial state. Those factors are also a threat to external validity as we cannot make any guarantee for models outside this scope.

\section{Related work}
\label{sec:related}

\paragraph*{CSP influences}
Failure traces are usually defined regarding linear transition systems (LTS). However, significant focus and practical development is on the CSP~\cite{Hoare1985} language. 
%
Many works by Wehrheim~\cite{Wehrheim2000,Wehrheim2003,Wehrheim2005,Wehrheim2006} cover the effect trace and failure divergence refinement has on behavioral properties within the context of CSP. Our work draws inspiration from this, as reasoning about traces can be more appealing than reasoning about POs (which is an alternative, as shown later). Traces represent what a system can do, so when \Cref{alg:algB} produces counterexamples, those may be easier to comprehend than failing proof.

As B is the direct predecessor of Event-B, the contribution of Dunne and Conroy~\cite{Dunne2005} is noteworthy as it provides POs for B such that CSP semantics like failure-divergence hold. Hallerstede~\cite{Hallerstede2011}, in his work, describes how CSP failure divergence refinement corresponds to deadlock freeness proofs of Event-B. In four contributions, Schneider et al.~\cite{Schneider2010,Schneider2011a,Schneider2011,Schneider2014} describe how CSP semantics relate to Event-B refinement. The main contribution here is a coupling of Event-B $\|$ CSP, allowing us to observe Event-B behavior regarding CSP traces. From this, the authors make further observations of the relationship between the Event-B refinement strategy and traces in CSP. 

These contributions aim to relate trace semantics to (Event-)B POs. Our contribution differs as we aim to add failure divergences semantics to Event-B without relating it to POs. This nonrelation, for now, is by choice, as we focus on getting a push-button tool ready that can directly assist in modeling chores like validating (trace) liveness properties for refinements without having to think about any proofs. Furthermore, in contrast to proofs, we can produce counter-examples, thus helping to debug models. 




\paragraph*{LT properties in Event-B}
 \label{subsec:lt_eventb}
Several significant related works aim to preserve LT(L) properties during Event-B refinement. Hoang and Abrial~\cite{Hoang2011} relate liveness properties in the context of LTL with new Event-B proof obligations, i.e., one could formulate LTL properties for the price of additional proofs to be discharged for every new property. Integrating LTL properties this way makes them accessible to the refinement mechanism. However, not all LTL property types are covered, and the new POs never became part of the standard Event-B POs or the Rodin toolset. Hoang et al.~\cite{Hoang2016} and Zhu et al.~\cite{Zhu2023} extend the findings to all LTL properties and their preservation during refinement. Rivi{\`e}re et al.~\cite{Riviere2023} proposed reasoning of liveness properties for reflexive Event-B; however, to our knowledge, it did not yet tackle the challenge of refinement.

\section{Conclusion and future work}
\label{sec:conclusion}
This work introduces failure divergence refinement for Event-B, enabling preservation assumptions about behavioral properties during refinement. Modelers can conduct the refinement check push-button through the algorithm and tool presented, contrasting previous methods requiring additional proofs. This offers swift, early feedback, facilitating the trickle-down of validation results along refinement chains without redoing validation. 

In the future, we aim to integrate the tool into ProB2-Ui~\cite{Bendisposto2021} for accessibility. As discussed in \Cref{sec:formalization}, an extension to LTL (or even CTL) is possible, and we plan to explore this. Furthermore, investigating readiness refinement/bisimulation (cf.~\cite{DeRoever1998}) is an intriguing avenue. Eshuis and Fokkinga~\cite{Eshuis2002} have observed a strong resemblance between failure divergence refinement and bisimulation for LTS, which warrants further exploration.


\bibliographystyle{splncs04.bst}
\bibliography{bib.bib}

\begin{thebibliography}{10}
\providecommand{\url}[1]{\texttt{#1}}
\providecommand{\urlprefix}{URL }
\providecommand{\doi}[1]{https://doi.org/#1}

\bibitem{Abrial2010}
Abrial, J.R.: Modeling in Event-B: System and Software Engineering. Cambridge University Press (2010)

\bibitem{Abrial2010a}
Abrial, J.R., Butler, M., Hallerstede, S., Hoang, T.S., Mehta, F., Voisin, L.: {R}odin: An open toolset for modelling and reasoning in {E}vent-{B}. International Journal on Software Tools for Technology Transfer  \textbf{12}(6),  447–466 (Nov 2010)

\bibitem{Bendisposto2021}
Bendisposto, J., Gele{\ss}us, D., Jansing, Y., Leuschel, M., P{\"u}tz, A., Vu, F., Werth, M.: {ProB2-UI}: A java-based user interface for prob. In: Lluch~Lafuente, A., Mavridou, A. (eds.) Formal Methods for Industrial Critical Systems. pp. 193--201. Springer International Publishing, Cham (2021)

\bibitem{DeRoever1998}
De~Roever, W.P., Engelhardt, K.: Data refinement: model-oriented proof methods and their comparison. Cambridge University Press, Cambridge, United Kingdom (1998)

\bibitem{Derrick2018}
Derrick, J., Boiten, E.: Refinement: Semantics, Languages and Applications. Springer, Cham (2018)

\bibitem{Dunne2005}
Dunne, S., Conroy, S.: Process refinement in {B}. In: Treharne, H., King, S., Henson, M., Schneider, S. (eds.) ZB 2005: Formal Specification and Development in Z and B. pp. 45--64. Springer Berlin Heidelberg, Berlin, Heidelberg (2005)

\bibitem{Eshuis2002}
Eshuis, R., Fokkinga, M.M.: Comparing refinements for failure and bisimulation semantics. Fundamenta Informaticae  \textbf{52}(4),  297--321 (2002)

\bibitem{Hallerstede2011}
Hallerstede, S.: On the purpose of {Event-B} proof obligations. Formal Aspects of Computing  \textbf{23},  133--150 (2011)

\bibitem{Hoang2013}
Hoang, T.S.: An introduction to the Event-B modelling method, pp. 211--236. Springer, Heidelberg (2013)

\bibitem{Hoang2011}
Hoang, T.S., Abrial, J.R.: Reasoning about liveness properties in {Event-B}. In: Qin, S., Qiu, Z. (eds.) Formal Methods and Software Engineering. pp. 456--471. Springer Berlin Heidelberg, Berlin, Heidelberg (2011)

\bibitem{Hoang2016}
Hoang, T.S., Schneider, S., Treharne, H., Williams, D.M.: Foundations for using linear temporal logic in event-b refinement. Formal Aspects of Computing  \textbf{28},  909--935 (2016)

\bibitem{Hoang2016a}
Hoang, T.S., Snook, C., Ladenberger, L., Butler, M.: Validating the requirements and design of a hemodialysis machine using {iUML-B}, {BMotion} studio, and co-simulation. In: Butler, M., Schewe, K.D., Mashkoor, A., Biro, M. (eds.) Abstract State Machines, Alloy, B, TLA, VDM, and Z. pp. 360--375. Springer International Publishing, Cham (2016)

\bibitem{Hoare1985}
Hoare, C.A.R., et~al.: Communicating sequential processes, vol.~178. Prentice-hall Englewood Cliffs (1985)

\bibitem{Leuschel2003}
Leuschel, M., Butler, M.: {ProB}: A model checker for {B}. In: Araki, K., Gnesi, S., Mandrioli, D. (eds.) FME 2003: Formal Methods. pp. 855--874. Springer Berlin Heidelberg, Berlin, Heidelberg (2003)

\bibitem{Leuschel2005}
Leuschel, M., Butler, M.: Automatic refinement checking for {B}. In: Lau, K.K., Banach, R. (eds.) Formal Methods and Software Engineering. pp. 345--359. Springer Berlin Heidelberg, Berlin, Heidelberg (2005)

\bibitem{Mashkoor2016}
Mashkoor, A.: The hemodialysis machine case study. In: Butler, M., Schewe, K.D., Mashkoor, A., Biro, M. (eds.) Abstract State Machines, Alloy, B, TLA, VDM, and Z. pp. 329--343. Springer International Publishing, Cham (2016)

\bibitem{mashkoor21a}
Mashkoor, A., Leuschel, M., Egyed, A.: Validation obligations: {A} novel approach to check compliance between requirements and their formal specification. In: 43rd {IEEE/ACM} International Conference on Software Engineering: New Ideas and Emerging Results, {ICSE} {(NIER)} 2021, Madrid, Spain, May 25-28, 2021. pp.~1--5. {IEEE} (2021). \doi{10.1109/ICSE-NIER52604.2021.00009}, \url{https://doi.org/10.1109/ICSE-NIER52604.2021.00009}

\bibitem{Riviere2023}
Rivi{\`e}re, P., Singh, N.K., A{\"i}t-Ameur, Y., Dupont, G.: Formalising liveness properties in {Event-B} with the reflexive {EB4EB} framework. In: Rozier, K.Y., Chaudhuri, S. (eds.) NASA Formal Methods. pp. 312--331. Springer Nature Switzerland, Cham (2023)

\bibitem{Roscoe1998}
Roscoe, A.: The theory and practice of concurrency  (1998)

\bibitem{Schneider2010}
Schneider, S., Treharne, H., Wehrheim, H.: A {CSP} approach to control in {Event-B}. In: M{\'e}ry, D., Merz, S. (eds.) Integrated Formal Methods. pp. 260--274. Springer Berlin Heidelberg, Berlin, Heidelberg (2010)

\bibitem{Schneider2011a}
Schneider, S., Treharne, H., Wehrheim, H.: Bounded retransmission in {Event-B} || {CSP}: a case study. Electronic Notes in Theoretical Computer Science  \textbf{280},  69--80 (2011). \doi{https://doi.org/10.1016/j.entcs.2011.11.019}, \url{https://www.sciencedirect.com/science/article/pii/S157106611100168X}, proceedings of the B 2011 Workshop, a satellite event of the 17th International Symposium on Formal Methods (FM 2011)

\bibitem{Schneider2011}
Schneider, S., Treharne, H., Wehrheim, H.: A {CSP} account of {Event-B} refinement (2011)

\bibitem{Schneider2014}
Schneider, S., Treharne, H., Wehrheim, H.: The behavioural semantics of {Event-B} refinement. Formal aspects of computing  \textbf{26},  251--280 (2014)

\bibitem{Stock2023}
Stock, S., Vu, F., Gele{\ss}us, D., Leuschel, M., Mashkoor, A., Egyed, A.: Validation by abstraction and refinement. In: Gl{\"a}sser, U., Creissac~Campos, J., M{\'e}ry, D., Palanque, P. (eds.) Rigorous State-Based Methods. pp. 160--178. Springer Nature Switzerland, Cham (2023)

\bibitem{Wehrheim2000}
Wehrheim, H.: Behavioural subtyping and property preservation. In: Smith, S.F., Talcott, C.L. (eds.) Formal Methods for Open Object-Based Distributed Systems IV. pp. 213--231. Springer US, Boston, MA (2000)

\bibitem{Wehrheim2003}
Wehrheim, H.: Behavioral subtyping relations for active objects. Formal Methods in System Design  \textbf{23},  143--170 (2003)

\bibitem{Wehrheim2005}
Wehrheim, H.: {Refinement and Consistency in Multiview Models}. In: Bezivin, J., Heckel, R. (eds.) Language Engineering for Model-Driven Software Development. Dagstuhl Seminar Proceedings (DagSemProc), vol.~4101, pp. 1--11. Schloss Dagstuhl -- Leibniz-Zentrum f{\"u}r Informatik (2005). \doi{10.4230/DagSemProc.04101.13}, \url{https://drops.dagstuhl.de/opus/volltexte/2005/19}

\bibitem{Wehrheim2006}
Wehrheim, H.: Refinement and consistency in component models with multiple views. In: Reussner, R.H., Stafford, J.A., Szyperski, C.A. (eds.) Architecting Systems with Trustworthy Components. pp. 84--102. Springer (2006)

\bibitem{Wright2010}
Wright, H.K., Kim, M., Perry, D.E.: Validity concerns in software engineering research. In: Proceedings of the FSE/SDP Workshop on Future of Software Engineering Research. p. 411–414. FoSER '10, ACM (2010). \doi{10.1145/1882362.1882446}, \url{https://doi.org/10.1145/1882362.1882446}

\bibitem{Zhu2019}
Zhu, C., Butler, M., Cirstea, C.: Towards refinement semantics of real-time trigger-response properties in {Event-B}. In: 2019 International Symposium on Theoretical Aspects of Software Engineering (TASE). pp.~1--8. IEEE (July 2019). \doi{10.1109/TASE.2019.00-26}

\bibitem{Zhu2023}
Zhu, C., Butler, M., Cirstea, C., Hoang, T.S.: A fairness-based refinement strategy to transform liveness properties in {Event-B} models. Science of Computer Programming  \textbf{225},  102907 (2023). \doi{https://doi.org/10.1016/j.scico.2022.102907}, \url{https://www.sciencedirect.com/science/article/pii/S016764232200140X}

\end{thebibliography}

\newpage
\appendix
\section{Proofs}
\label{sec:proofs}

\begin{proof}
[Proof of \Cref{def:prop:convergence_induces_stable_states}]
   For the proof, we introduce the concatenation operator $^\frown$ for traces, e.g., $\sigma^\frown\pi$ is defined if the last state of $\sigma$ is equal to the first state of $\pi$. (By contradiction) Assume that  $\Pi \neq \emptyset$ but $\nexists \pi \cdot \pi \in \Pi$ such that $last(\pi) \in stable_C$. Then any $\pi \in \Pi$ can be extended by an event $i \in \mathcal{N}_C$ such that $\tau(\pi^\frown i) = \tau(\pi)$. However, by our assumption, the trace of $\pi^\frown i$ has also not a stable end state. Therefore, we  extend the trace further with an event $j \in \mathcal{N}_C$ such that $\tau(\pi^\frown i^\frown j) = \tau(\pi^\frown i) = \tau(\pi)$. Again, $\pi^\frown i^\frown j$ is unstable in the last state. This can be continued indefinitely. As our state space is finite, we eventually will encounter a state we have already seen, which is against the definition of divergence freeness.
\end{proof}

\begin{proof} [Proof of \Cref{def:prop:mainProp}]
    (By contradiction) \textbf{Case 1:} $\sigma \in traces(A)$ but $\sigma \notin \tau(traces(C))$. Then there must exist a maximum depth in the transition system of A such that we can find a common prefix of $\sigma$ called $\pi$ such that $\pi \in traces(A)$ and $\pi \in \tau(traces(C))$. The corresponding trace in C would then be  $\pi' \in traces(C)$ with $last(\pi) \in stable_C$\footnote{As we demand the absence of divergence we can assume a stable state with \Cref{def:prop:convergence_induces_stable_states}.} and $X = \mathit{refusals}(last(\pi)$). 
    As this is the longest common prefix, the following would hold $(\tau(\pi'), \mathit{AbsRefusal}(X)) \notin failure(A)$; however, as $A \sqsubseteq_{fd} C$, this is a contradiction. \newline
     \textbf{Case 2:} $\sigma \in traces(A)_\Delta$ but $\sigma \notin \tau(traces(C)_\Delta)$. A finite longest common prefix must be within a finite state space. After finding this prefix, we reuse the proof from Case 1. \newline
     \textbf{Note} that the divergence check fulfills the additional purpose that finite trace cannot suddenly become infinite during refinement. Analogously, the check for failure equivalence prevents traces from becoming prematurely deadlocked.
\end{proof}

\end{document}